\documentclass[preprint,showpacs,showkeys,secnumarabic]{revtex4}
\usepackage{amsmath}

\newcommand{\pni}{\par\noindent}
\begin{document}
\title{Cosmological solutions in the brane-bulk system}
\author {M. La Camera}
\email{lacamera@ge.infn.it} 
\affiliation {Department of Physics and INFN - University of 
Genoa\\Via Dodecaneso 33, 16146 Genoa, Italy} 
\begin{abstract}
In this work we find  cosmological solutions in the 
brane-bulk system starting from a 5-D line element which is a 
simple extension, for cosmological applications, of the 
pioneering Randall-Sundrum line element.  From the knowledge of 
the bulk metric, assumed to have the form of plane waves 
propagating in the fifth dimension, we  solve the  corresponding 
4-D Einstein equations on the brane with a well defined 
energy-momentum tensor. 
\end{abstract} 
\pacs{04.20.Jb; 04.50.+h }
\keywords{Brane Theory; General Relativity }
\maketitle  
\section{I\lowercase{ntroduction}}
Brane cosmologies are often studied in the framework of
five-dimensional (5-D) Einstein equations in the bulk. The 
effective 4-D gravitational  equations in the brane without 
curvature correction terms were first obtained by Shiromizu, 
Maeda and Sasaki [1]. However, even taking more generalized 
gravitational actions, the derived 4-D Einstein equations do not 
in general form a closed system due to the presence of a Weyl 
term which can only be specified in terms of the bulk metric, so 
other equations are to be written down and different procedures 
arise in splitting the non-Einsteinian terms between bulk and 
brane [2]. Because the specific form of a solution in the 
bulk is in general rather cumbersome due to the number of terms 
and parameters in the equations, we shall consider, as a guide 
for further work, a model simple enough to allow obtaining not 
trivial exact solutions. This paper is organised as follows. In 
the next Section we summarize the effective brane equations 
obtained by Kofinas [3]  when the intrinsic curvature scalar
$^{(4)}R$ is included in the brane action. In Section 3 we 
transform the static line element proposed by Randall and Sundrum
[4] into a dynamical one containing a three-space with 
constant curvature and obtain, using the 5-D Einstein equations, 
the corresponding energy-momentum tensor in the bulk. In Section 
4 we find the related bulk metric in  coordinate systems 
commonly used in cosmological applications. Solutions in the 
brane, assumed infinitely thin and $\mathbf{Z_2}$ symmetric in 
the bulk, are found in Section 5. Finally, conclusions are given 
in Section 6. 
\section{B\lowercase{raneworld} E\lowercase{instein} 
\lowercase{field equations}} 
In this section we recall the  effective brane equations obtained
by  Kofinas [3], which we shall use in the following, giving
a brief account of their derivation. Once we have solved the 
equations in the bulk, the form of the induced equations will 
allow us finding brane solutions following the methods of General
Relativity with a well defined energy-momentum tensor. The 
starting point in [3] is a three-dimensional brane $\Sigma$ 
embedded in a five-dimensional spacetime $M$. For convenience a 
coordinate $y$ is chosen such that the hypersurface $y = 0$ 
coincides with the brane. The total action for the system is 
taken to be 
\begin{eqnarray} 
\mathcal{S} && = \dfrac{1}{2\,\kappa_{5}^2}\, \int_M\, 
\sqrt{-\,^{(5)}g} \,(^{(5)}R - 2\Lambda_5)\,d^5x + 
\dfrac{1}{2\,\kappa_{4}^2} \,\int_\Sigma\, \sqrt{-\,^{(4)}g} 
\,(^{(4)}R - 2\Lambda_4)\,d^4x \nonumber\\ && \,\, +  \int_M\, 
\sqrt{-\,^{(5)}g} \,L_5^{mat}\,d^5x + \int_\Sigma\, 
\sqrt{-\,^{(4)}g} \,L_4^{mat}\,d^4x 
\end{eqnarray} 
Bulk indices will be denoted by capital Latin letters and brane 
indices by lower Greek letters. Varying (1) with respect to the  
bulk metric $g_{AB}$ one obtains the equations
\begin{equation}
^{(5)}G_A^B = - \Lambda_5 \delta_A^B + \kappa_5^2\, (^{(5)}T_A^B 
+ ^{(loc)}T_A^B \delta(y)) 
\end{equation}
where
\begin{equation}
^{(loc)} T_A^B = -\, 
\dfrac{1}{\kappa_4^2}\,\sqrt{\dfrac{-\,^{(4)}g}{-\,^{(5)}g}}\, 
(^{(4)}G_A^B - \kappa_4^2\,{^{(4)}}T_A^B + \Lambda_4h_A^B)
\end{equation}
is the localized energy-momentum tensor of the brane. 
$^{(5)}G_{AB}$ and $^{(4)}G_{AB}$ denote the Einstein tensors 
constructed from the bulk and the brane metrics respectively, 
while the tensor $h_{AB} = g_{AB} -n_An_B$ is the induced metric 
on the hypersurfaces $y$ = constant, with $n^A$ the normal unit 
vector on these
\begin{equation}
n^A = \dfrac{\delta_5^A}{\Phi}, \hspace{0.5in} n_A = 
(0,0,0,0,\Phi) 
\end{equation}
The way the coordinate $y$ has been chosen allows to write the  
five-dimensional line element, at least in the neighborood of the
brane, as 
\begin{equation}
dS^2 = g_{AB}\, dx^A dx^B = g_{\mu\nu}\, dx^\mu dx^\nu + 
\Phi^2dy^2 
\end{equation}
Using the methods of canonical analysis [5] the Einstein 
eqs. (2) in the bulk  are split into the following sets of 
equations 
\begin{subequations}
\begin{eqnarray} 
K_{\mu;\nu}^\nu -\, K_{;\mu}  &&=\, \kappa_5^2\, 
\Phi\,^{(5)}T_\mu^y \label{equation20}\\
K_\nu^\mu K_\mu^\nu - K^2 +\, {^{(4)}R}  &&=\, 2\, (\Lambda_5 
-\, \kappa_5^2\,{^{(5)}}T_y^y) \label{equation21}\\ 
\hspace{-0.3in} \dfrac{\partial K_\nu^\mu}{\partial y} + 
\Phi\,K\,K_\nu^\mu -\, \Phi\,^{(4)}R_\nu^\mu + 
g^{\mu\lambda}\,\Phi_{;\lambda\nu} &&=\, 
-\,\kappa_5^2\,\Phi\,\left(^{(loc)}T_\nu^\mu -\, 
\dfrac{1}{3}\,^{(loc)} T\,\delta_\nu^\mu\right)\,\delta(y) 
-\,\kappa_5^2\,\Phi\,^{(5)} T_\nu^\mu\nonumber\\ 
&&\quad\, -\, \kappa_5^2\,\Phi\,^{(5)}T_\nu^\mu 
+\dfrac{1}{3}\,\Phi\,(\kappa_5^2\,{^{(5)}T} -\, 
2\Lambda_5)\,\delta_\nu^\mu \label{equation22} 
\end{eqnarray} 
\end{subequations}
where $K_{\mu\nu}$ is the extrinsic curvature of the 
hypersurfaces $y$ = constant:
\begin{equation} 
K_{\mu\nu} = \dfrac{1}{2\,\Phi}\,\dfrac{\partial 
g_{\mu\nu}}{\partial y}, \quad K_{A y} = 0 
\end{equation} 
The Israel conditions [6] for the singular part in  
eqs. (6c) are
\begin{equation}
[K_\nu^\mu] = -\kappa_5^2\,\Phi_0\,\left(^{(loc)}T_\nu^\mu -\, 
\dfrac{1}{3}\,^{(loc)} T\,\delta_\nu^\mu\right) 
\end{equation}
where the bracket means discontinuity of the quantity across $y =
0$ and $\Phi_0 = \Phi(y=0)$. Hereafter, considering a 
$\mathbf{Z_2}$ symmetry on reflection around the brane, eqs. (3) 
becomes
\begin{equation}
^{(4)}G_\nu^\mu = - \Lambda_4\,\delta_\nu^\mu + \kappa_4^2\, 
{^{(4)}T_\nu^\mu}  + \dfrac{2}{r_c}\,(\overline{K}_\nu^\mu
- \overline{K}\delta_\nu^\mu)
\end{equation}
where \,$\overline{K}_\nu^\mu = K_\nu^\mu(y=0^+) = 
-\,K_\nu^\mu(y=0^-)$\, and \,$r_c = 
\kappa_5^2/\kappa_4^2$\, is a distance scale.
Equations (9) assume the form of usual Einstein equations with 
the energy-momentum tensor splitted into the common brane 
energy-momentum tensor plus additional terms which are 
all multiplied by $1/r_c$. The tensor  $^{(4)}T_\nu^\mu$ 
satisfies the usual conservation law $^{(4)}T_{\nu ;\mu}^\mu = 0$
provided $^{(5)}T_\mu^y =0$, which means no exchange of 
energy between brane and bulk. The quantities 
$\overline{K}_\nu^\mu$ are still undetermined, however additional
information can be obtained from the geometrical identity 
\begin{equation} ^{(4)}R_{BCD}^A =\, 
^{(5)}R_{NKL}^M\,h_M^A\,h_B^N\,h_C^K\,h_D^L + (K_C^A\,K_{BD} - 
K_D^A\,K_{BC}) \end{equation} 
Taking suitable contractions from 
the above relation it is possible to construct the four and 
five-dimensional Einstein tensors and, making use of the bulk 
Einstein equations, to get finally the parallel to the brane 
equations 
\begin{eqnarray}
^{(4)}G_\nu^\mu = && -\,\dfrac{1}{2}\,\Lambda_5 \delta_\nu^\mu 
+ \dfrac{2}{3}\,\kappa_5^2\, \left({}^{(5)}\overline{T}_\nu^\mu +
\left({}^{(5)}\overline{T}_y^y -  
\dfrac{1}{4}\,{}^{(5)}\overline{T}\right) \delta_\nu^\mu  \right)
\nonumber \\ 
&& +\left(\overline{K}\,\overline{K}_\nu^\mu - 
\overline{K}_\lambda^\mu\, \overline{K}_\nu^\lambda \right) + 
\dfrac{1}{2}\, \left(\overline{K}_\lambda^\kappa\, 
\overline{K}_\kappa^\lambda - K^2\right) \delta_\nu^\mu  - 
g^{\kappa\mu}\,{}^{(5)}\overline{C}_{\kappa y \nu}^y 
\end{eqnarray}
Here ${}^{(5)}C_{\kappa y \nu}^y$ is the ``electric'' part of 
the bulk Weyl tensor, while $\overline{T}$ and $\overline{C}$ are
the limiting values of those quantities at $y = 0^+$ or $0^-$.
Now, equating the right-hand sides of the independent 
eqs. (9) and (11), one gets an algebraic equation  for 
$\overline{K}_\nu^\mu$ which can be substituted in (9) once the 
equation has been solved. However the system of Einstein 
equations for the brane metric so obtained is not, in general, 
closed due to the presence of the Weyl term, so one has to solve 
the Einstein equations in the bulk in order to determine the Weyl
tensor on the brane. This one will be the method followed in this
paper. A different approach, which instead does not assume a bulk
geometry, starts from deducing a brane dynamics and then searches
for a bulk geometry in which the brane can consist its boundary~ 
[7].
\section{T\lowercase{he model}} 
Following a possibility suggested in [8] we shall consider 
5-D Einstein equations strictly in the bulk, i.e. without the 
brane energy-momentum tensor with its delta distribution,  so 
when searching, later on, for solutions in the brane we 
shall have to take limiting values of bulk quantities as 
discussed in the context of eqs. (11). As a simplifying 
device for dealing with these equations, we select a 
five-dimensional line element with reasonable physical 
properties, calculate  the corresponding energy-momentum tensor 
$^{(5)}T_A^B$ and, once solved eqs. (11) for the brane 
metric, we can obtain, from eqs. (9), the 4-D cosmological 
constant $\Lambda_4$ and the effective energy- momentum tensor on
the brane. More in detail, let us consider the static 
Randall-Sundrum line element 
\begin{equation} 
dS^2 = e^{-2 \kappa y}\,\left( A^2\, d\sigma_k^2
-dt^2 \right)+ dy^2 
\end{equation}
containing a three-space with constant curvature.
Here $\kappa$ and $A$ are, respectively, the constant scale 
factors for the extra dimension $y$ and for the ordinary 
three-space and $d\sigma_k^2$ is the line element on maximally 
symmetric three-spaces with curvature index $k = +1, 0, -1$:
\begin{equation}
d\sigma_k^2 = \dfrac{dr^2}{1-k r^2} + r^2  
(d\vartheta^2+\sin^2\vartheta d\varphi^2)
\end{equation}
Since  our purpose is to describe the time evolution on the 
braneworld,  we need to transform the static bulk solution (12) 
into a dynamical one. We follow the procedure used in 
[9,10,11] where dynamical solutions are derived from the 
static Randall-Sundrum metric by generalized boosts along the 
fifth dimension. Applied to the actual case, we write the 
required transformations as
\begin{equation} 
\begin{cases}
t = & \dfrac{\dfrac{1-e^{-\widetilde{\chi}\, 
\overline{t}}}{\chi}-\dfrac{\chi} 
{\kappa^2}e^{\widetilde{\kappa}\, 
\overline{y}}}{\sqrt{1-\dfrac{\chi^2}{\kappa^2}}} \\ {}& \\ 
e^{\kappa\, y} = & \dfrac{e^{\widetilde{\kappa}\, \overline{y}}+ 
e^{- \widetilde{\chi}\, \overline{t}}-1} 
{\sqrt{1-\dfrac{\chi^2}{\kappa^2}}} 
\end{cases} 
\end{equation} 
where \, $\widetilde{\kappa} = \dfrac{\kappa}{\sqrt{1 - 
\dfrac{\chi^2}{\kappa^2}}}$,\,\,
$\widetilde{\chi} = \dfrac{\chi}{\sqrt{1 - 
\dfrac{\chi^2}{\kappa^2}}}$\,\, 
and $\chi$ is a constant responsible of the boost in the  
($t,y$) spacetime.  \pni As a result the metric (12) 
becomes, dropping the bar: 
\begin{equation}
dS^2 = \dfrac{1}{(e^{-\widetilde{\chi}\, t}+ e^{\widetilde{\kappa}\, 
y}-1)^2}\, \left( \widetilde{A}^2\, d\sigma_k^2-e^{-2 \widetilde{\chi} 
\, t}\,dt^2 + e^{2 \widetilde{\kappa}\, y}\, dy^2 \right) 
\end{equation}
with\, $\widetilde{A}^2 = (1-\dfrac{\chi^2}{\kappa^2})\, A^2$. \pni
Having  specified  the metric components and consequently the 
components of the Einstein tensor, we have
\begin{subequations} 
\begin{eqnarray} 
&& G_r^r = G_\vartheta^\vartheta = G_\varphi^\varphi  = -\,6\, 
(\widetilde{\chi}^2 -  \widetilde{\kappa}^2) -\dfrac{k}{a^2(t,y)} 
\label{equationa} \\  
&& G_t^t = G_y^y =  -\,6\, (\widetilde{\chi}^2 - \widetilde{\kappa}^2) - 
\dfrac{3\, k}{a^2(t,y)} \label{equationb} 
\end{eqnarray}
\end{subequations} 
where 
\begin{equation}
a(t,y) = \dfrac{\widetilde{A}}{e^{-\widetilde{\chi}\, t}+ 
e^{\widetilde{\kappa}\, y}-1} 
\end{equation} 
By comparison with (16) and taking into account that $G_t^r$ 
and $G_t^y$ are both equal to zero, we obtain 
\begin{equation}
\Lambda_5 = 6\,(\widetilde{\chi}^2 -  \widetilde{\kappa}^2) = -\, 
6\,\kappa^2 , \qquad ^{(5)}T_A^B = 
\textrm{diag}(p,p,p,-\,\rho,p_\perp) \end{equation}
with
\begin{equation}
\kappa_5^2\,p = -\,\dfrac{k}{a^2(t,y)}, \quad \kappa_5^2\,\rho
= -\,\kappa_5^2\,p_\perp =  \dfrac{3\,k}{a^2(t,y)} 
\end{equation}
It is worth noticing that the bulk sources contain, 
besides the cosmological term which can be interpreted as 
proportional  to the pressure and to the density of vacuum 
fluctuations [12],  terms proportional to the curvature 
index $k$ which call in mind, even if $k$ and $a^2(t,y)$  refer 
to different spaces, a relationship with the ``pressure'' and 
``density'' coming from the curvature of the universe [13]. 
The line element (15) is a particular realization of the metric 
\begin{equation}
dS^2 = a^2(t,y)\,d\sigma_k^2 - n^2(t,y)\,dt^2 +  
\Phi^2(t,y)\,dy^2
\end{equation}
commonly used in cosmological applications, so it is worth  
obtaining  the above metric coefficients   solving  Einstein
with a generic cosmological constant $\Lambda_5$ 
and the tensor $^{(5)}T_A^B$ whose  components are given in (19).
\section{S\lowercase{olutions in the bulk}} 
In the coordinate system (20) the non-vanishing components of the 
Einstein tensor $G_A^B$ are 
\begin{subequations}
\begin{eqnarray}  \hspace{-1in}
G_r^r = G_\vartheta^\vartheta = G_\varphi^\varphi = 
&&-\,\dfrac{1}{n^2}\left[\dfrac{\ddot{\Phi}}{\Phi} + \dfrac{2 
\ddot{a}}{a} + \dfrac{\dot{\Phi}}{\Phi}\,
\left(\dfrac{2\dot{a}}{a} - \dfrac{\dot{n}}{n}\right) + 
\dfrac{\dot{a}}{a}\,\left(\dfrac{\dot{a}}{a} - \dfrac{2 
\dot{n}}{n}\right)\right]  \nonumber \\
&&  +\,\dfrac{1}{\Phi^2}\,\left[\dfrac{2a''}{a} + 
\dfrac{n''}{n} +\dfrac{a'}{a}\,\left(\frac{a'}{a} + 
\dfrac{2n'}{n}\right) - \dfrac{\Phi'}{\Phi}\,\left(\dfrac{2a'}{a}
+ \dfrac{n'}{n}\right)\right] - \dfrac{k}{a^2} 
\label{equationc}\\
G_t^t = &&  -\,\dfrac{3}{n^2}\,\left(\dfrac{\dot{a}^2}{a^2}+ 
\dfrac{\dot{a}\dot{\Phi}}{a\Phi}\right) + 
\dfrac{3}{\Phi^2}\, \left(\dfrac{a''}{a} + 
\dfrac{{a'}^2}{a^2} - \dfrac{a'\Phi'}{a\Phi}\right) - 
\dfrac{3k}{a^2} \label{equationd}\\
G_y^y =&&   -\,\dfrac{3}{n^2}\,\left(\dfrac{\ddot{a}}{a} + 
\dfrac{\dot{a}^2}{a^2} - \dfrac{\dot{a}\dot{n}}{an}\right) + 
\dfrac{3}{\Phi^2}\,\left(\dfrac{{a'}^2}{a^2} + 
\dfrac{a'n'}{an}\right) - \dfrac{3k}{a^2} \label{equatione}\\
G_y^t =&& 
 -\,\dfrac{3}{\Phi^2}\,\left(\dfrac{\dot{a}'}{a} - 
\dfrac{\dot{a}n'}{an} - \dfrac{a'\dot{\Phi}}{a\Phi}\right) 
\label{equationf} 
\end{eqnarray}
\end{subequations}
Here a dot and a prime denote partial derivatives with respect to
$t$ and $y$, respectively. The specific form of the solution in 
the bulk is in general rather cumbersome, but some 
simplifications can arise if one makes the assumption of plane 
waves solutions, i.e. if one assumes that the metric coefficients
are functions of the argument $u = (t - \lambda\,y)$ or of
$v = (t + \lambda\,y)$ [14,15,16,17]. 
In the following we concentrate on solutions which depend on  
$(t - \lambda\,y)$: 
\begin{equation}
a = a(t - \lambda\,y), \quad n=n(t - \lambda\,y), \quad \Phi = 
\Phi(t - \lambda\,y) 
\end{equation}
From the equation $G_y^t = 0$ we get
\begin{equation}
\dfrac{\overset{\ast}{a}}{n\Phi} = \text{constant}
\end{equation}
where the superscribed asterisk $\overset{\ast}{\frac{}{}}$ 
denotes derivative with respect to $u$. Now considering the 
history of the three-brane as described by a point trajectory in 
the ($t,y$) spacetime it is possible to introduce a 
Gaussian normal coordinate system where  $\Phi = 1$. 
Another ansatz where $n = 1$ is made in the 
literature [18,19] so we shall separately consider these two
additional assumptions to solve the Einstein equations in the 
bulk. \pni Let us begin with the choice $n = 1$. One has from 
(23)
\begin{equation}
\Phi = \dfrac{\overset{\ast}{a}}{\beta} 
\end{equation}
with $\beta$ a constant.
Now $G_t^t = G_y^y$ and the Einstein equations reduce to
\begin{subequations}
\begin{eqnarray}
\dfrac{a^2\, \overset{\ast\ast\ast}{a}}{\overset{\ast}{a}} + 4\, 
a\, \overset{\ast\ast}{a} + {\overset{\ast}{a}}^2 && = 
\Lambda_5\, a^2 + \beta^2 \lambda^2 
\label{equation1} \\ a\, \overset{\ast\ast}{a} + 
{\overset{\ast}{a}}^2 && = \dfrac{\Lambda_5}{3}\, a^2 + 
\beta^2 \lambda^2 \label{equation2} 
\end{eqnarray}
\end{subequations}
Subtracting (25b) from (25a) we obtain an equation which 
is the first derivative of (25b) with respect to $u$, so both 
eqs. (25) are satisfied solving
\begin{equation}
\overset{\ast\ast}{(a^2)} = \dfrac{2}{3}\Lambda_5\, a^2 + 2\, 
\beta^2 \lambda^2
\end{equation}
The solution is
\begin{equation}
a^2 = c_1\,\sinh{\sqrt{\dfrac{2}{3}\,\Lambda_5}\,(t - 
\lambda\,y)} + c_2\,\cosh{\sqrt{\dfrac{2}{3}\,\Lambda_5}\,(t - 
\lambda\,y)} - 3\, \dfrac{\beta^2\lambda^2} 
{\Lambda_5}
\end{equation}
with $c_1$ and $c_2$ suitable constants. Requiring $a^2(0) = 0$ 
and putting for future simplifications $c_1 = 0$ we have 
\begin{equation}
a^2 = \dfrac{6\,\beta^2 \lambda^2}{ \Lambda_5}\, 
\sinh^2{\sqrt{\dfrac{\Lambda_5}{6}}\, (t - \lambda\,y)}
\end{equation}
The other metric coefficient is 
\begin{equation}
\Phi^2 = 
\lambda^2\,\cosh^2{\sqrt{\dfrac{\Lambda_5}{6}}\,(t - 
\lambda\,y)} 
\end{equation}
We notice that the choice $c_1 = 0$   implies 
$\overset{\ast}{(a^2)}(0) = 0$ and leaves  the possibility    
of taking either sign for the 5-D cosmological constant. \pni
When $\Phi = 1$ one has from (23) 
\begin{equation}    
n = \dfrac{\overset{\ast}{a}}{\alpha}
\end{equation}
with $\alpha$ a constant.
One has again $G_t^t = G_y^y$ and the Einstein equations reduce 
to 
\begin{subequations}
\begin{eqnarray}
\dfrac{a^2\, \overset{\ast\ast\ast}{a}}{\overset{\ast}{a}} + 4\, 
a\, \overset{\ast\ast}{a} + {\overset{\ast}{a}}^2 && = 
-\,\dfrac{\Lambda_5}{\lambda^2}\, a^2 + 
\dfrac{\alpha^2}{\lambda^2} \label{equation3} \\ 
a\, \overset{\ast\ast}{a} + {\overset{\ast}{a}}^2 && = 
-\,\dfrac{\Lambda_5}{3\, \lambda^2}\, a^2 + 
\dfrac{\alpha^2}{\lambda^2} \label{equation4} 
\end{eqnarray} 
\end{subequations}
Subtracting (31b) from (31a) we obtain an equation which 
is the first derivative of (31b) with respect to $u$, so both 
eqs. (31) are satisfied solving
\begin{equation}
\overset{\ast\ast}{(a^2)} = 
-\,\dfrac{2\,\Lambda_5}{3\,\lambda^2}\, a^2 + 
2\, \dfrac{\alpha^2}{\lambda^2} 
\end{equation}
The solution is
\begin{equation}
a^2 = 
c_1\,\sinh{\dfrac{1}{\lambda}\sqrt{-\,\dfrac{2}{3}\,
\Lambda_5}\,(t- \lambda\,y)} + c_2\,
\cosh{\dfrac{1}{\lambda}\sqrt{-\,\dfrac{2}{3}\,
\Lambda_5}\,(t - \lambda\,y)} - 
\dfrac{3\,\alpha^2}{\Lambda_5} 
\end{equation}
Requiring $a^2(0) = 0$ and putting for future simplifications 
$c_1 = 0$ we have
\begin{equation}
a^2 = -\,\dfrac{6\,\alpha^2}{\Lambda_5}\, 
\sinh^2{\dfrac{1}{\lambda}\sqrt{\dfrac{-\,\Lambda_5}{6}}\,
(t - \lambda\,y)} 
\end{equation}
The other metric coefficient is
\begin{equation}
n^2 = 
\dfrac{1}{\lambda^2}\,\cosh^2{\dfrac{1}{\lambda} 
\sqrt{\dfrac{-\,\Lambda_5}{6}}\, (t - \lambda\,y)} 
\end{equation} 
Here again the choice $c_1 = 0$ implies 
$\overset{\ast}{(a^2)}(0) = 0$ and leaves  the possibility of taking
either sign for the 5-D cosmological constant. When 
$\mathbf{Z_2}$ symmetry on reflection around the brane at $y = 0$
is assumed, the extra-coordinate $y$ in the above solutions must 
be replaced by $|y|$. 
\section{S\lowercase{olutions in the brane}}
The 4-D line element can be written as 
\begin{equation} 
ds^2 =\, a_0^2(t)\,d\sigma_k^2 - n_0^2(t)\,dt^2 
\end{equation} 
where the subscript ${}_0$ here does not mean ``calculated 
putting $y=0$ in the corresponding bulk quantities'' because 
those metric coefficients were found without taking into 
account the matter content of the brane, but means ``calculated 
solving Einstein equations (11) on the brane with 
$^{(4)}G_\nu^\mu$ derived from the line element (36)''. 
Before searching for plane wave solutions in the brane let us 
return, for completeness, to the case treated in Section 
3 with 5-D line element given in (15). The relevant Einstein 
equations obtained from (11), are 
\begin{subequations} 
\begin{eqnarray} \hspace{-0.5in}
^{(4)}G_r^r =\, ^{(4)}G_\vartheta^\vartheta =\, 
^{(4)}G_\varphi^\varphi =  -\,\dfrac{k}{a_0^2} -\, 
\dfrac{{\dot{a}_0}^2}{a_0^2 n_0^2} + 
\dfrac{2\,\dot{a}_0\dot{n}_0}{a_0 n_0^3} -\, 
\frac{2\,\ddot{a}_0}{a_0 n_0^2}\, = && -\,\dfrac{\Lambda_5}{2} 
-\,7\,\widetilde{\kappa}^2 - \dfrac{k\,e^{-2\,\widetilde{\chi} 
t}}{\widetilde{A}^2} \label{equation5} \\ 
^{(4)}G_t^t \,=-\,\dfrac{3\,k}{a_0^2} -\, 
\dfrac{3\,\dot{a}_0^2}{a_0^2 n_0^2}\, = && 
-\,\dfrac{\Lambda_5}{2} -\,7\,\widetilde{\kappa}^2 -\, 
\dfrac{3\,k\,e^{-2\,\widetilde{\chi} t}}{\widetilde{A}^2} 
\label{equation6} 
\end{eqnarray} 
\end{subequations} 
where $\Lambda_5$ is given in (18).\pni 
The solutions are: 
\begin{eqnarray}
a_0^2 = && \,c_1\, e^{2\, \widetilde{\chi}\, t}  \equiv   
\,\widetilde{A}^2\, e^{2\, \widetilde{\chi}\, t}\\
n_0^2 = && \dfrac{{\dot{a}_0}^2}{\dfrac{a_0^2}{3}\,
\left(\Lambda_4 + \kappa_4^2\,\rho\right)-\,k} = 1 
\end{eqnarray}
where $\Lambda_4$ and $\rho$, written explicitly below, are 
respectively the cosmological constant and the matter density 
in the brane. With the choice $c_1 = \widetilde{A}^2$ the brane 
metric can be obtained evaluating the corresponding bulk metric 
at $y = 0$. Comparing eqs. (37) and (9) we have defined 
the 4-D cosmological constant, including a possible contribution 
coming from the brane tension, as 
\begin{equation}
\Lambda_4 = \dfrac{\Lambda_5}{2} + 7\,\widetilde{\kappa}^2 =
\dfrac{4\,\kappa^2 + 3\,\chi^2}{1-\dfrac{\chi^2}{\kappa^2}}
\end{equation} 
The remaining terms are the components of the  effective 
energy-momentum tensor of the brane. This can be considered as 
the stress tensor of a perfect fluid at rest, because $G_{rt} = 
0$ in the frame given by (36), with pressure $p$  and density 
$\rho$ given by 
\begin{equation}
p = -\, \dfrac{1}{3}\,\rho, \quad \rho =  
\dfrac{3\,k}{\kappa_4^2\,a_0^2} 
\end{equation}
Here and in the following we shall attribute to a ``fluid'' also 
quantities proportional to the  curvature index $k$ and to the 
cosmological constant $\Lambda_4$; moreover all the fluids taken 
into account may cause violations of some of the energy 
conditions. Comparing the values of the Hubble constant obtained 
from its definition and from the Friedmann equation (37b)  
\begin{equation} 
H^2 = \left(\dfrac{\dot{a}_0}{a_0 n_0}\right)^2 
= \widetilde{\chi}^2 = \dfrac{\Lambda_4}{3} - \dfrac{k}{a_0^2} + 
\dfrac{\kappa_4^2}{3}\,\rho  = \dfrac{\Lambda_4}{3}
\end{equation}
we obtain $\widetilde{\chi} = H = \sqrt{\Lambda_4/3}$ for any 
value of the curvature index $k$  so  $a_0 \propto 
e^{H \, t}$  as in the de Sitter model, being now $t$ the proper 
time in the brane. Clearly the deceleration  parameter  $q = 
-\,(a_0\,\ddot{a}_0)/{\dot{a}}^2$ and the density parameter 
$\Omega_\Lambda = \Lambda_4/(3\,H^2)$ have the values $q = - 1$ 
and $\Omega_\Lambda = 1$.\pni
The next case we shall treat corresponds to the  choice $n = 1$ 
in the bulk. The relevant Einstein equations now are 
\begin{subequations} 
\begin{eqnarray} \hspace{-0.7in}
-\,\dfrac{k}{a_0^2} -\, \dfrac{{\dot{a}_0}^2}{a_0^2 n_0^2} + 
\dfrac{2\,\dot{a}_0\dot{n}_0}{a_0 n_0^3} -\, 
\frac{2\,\ddot{a}_0}{a_0 n_0^2}\, = &&  -\,\dfrac{\Lambda_5}{2} 
-\, \dfrac{2\,\Lambda_5}{3\, \sinh^2{\sqrt{\dfrac{\Lambda_5}
{6}}\,t}} -\, \dfrac{k}{\dfrac{6\,\beta^2 \lambda^2}{\Lambda_5}\,
\sinh^2{\sqrt{\dfrac{\Lambda_5}{6}}\,t}} 
\label{equation7} \\
-\,\dfrac{3\,k}{a_0^2} -\, \dfrac{3\,\dot{a}_0^2}{a_0^2 n_0^2}\, 
= && -\,\dfrac{\Lambda_5}{2} -\, \dfrac{\Lambda_5} 
{\sinh^2{\sqrt{\dfrac{\Lambda_5} {6}}\,t}} -\, 
\dfrac{3\,k}{\dfrac{6\,\beta^2 \lambda^2}{\Lambda_5}\, 
\sinh^2{\sqrt{\dfrac{\Lambda_5}{6}}\,t}} \label{equation8}
\end{eqnarray}
\end{subequations}
The solutions to the system (43) are
\begin{eqnarray}
a_0^2 = && c_1\, 
\left(\sinh^2{\sqrt{\dfrac{\Lambda_4}{3}}t}\right)^ 
{2-\,k/(k+\beta^2\lambda^2)} 
\\ n_0^2 = && 
\dfrac{{\dot{a}_0}^2}{\dfrac{a_0^2}{3}\,\left(\Lambda_4 + 
\kappa_4^2\,(\rho_1 + \rho_2) \right)-\,k}
\end{eqnarray}
Comparing eqs. (43) and (9), we have defined the 4-D 
cosmological constant as $\Lambda_4 = \dfrac{\Lambda_5}{2}$
and have considered the remaining terms as the superposition of 
two different fluids  with pressures and densities given by
\begin{eqnarray}   
p_1 =  -\,\dfrac{2}{3}\,\rho_1 , \quad 
\rho_1 = && \dfrac{2\,\Lambda_4} 
{\kappa_4^2\,\sinh^2{\sqrt{\dfrac{\Lambda_4} {3}}\,t}}\,\propto\,
a_0^{-\,2\,(k+\beta^2\lambda^2)/(k+2\,\beta^2\lambda^2)}  \\
p_2 = -\,\dfrac{1}{3}\,\rho_2, \quad \rho_2
= && \dfrac{k}{\dfrac{\beta^2 \lambda^2 
\kappa_4^2}{\Lambda_4}\, \sinh^2{\sqrt{\dfrac{\Lambda_4}{3}}\,t}}
\,\propto\, 
a_0^{-\,2\,(k+\beta^2\lambda^2)/(k+2\,\beta^2\lambda^2)} 
\end{eqnarray} 
From (43b) we obtain the Friedmann equation
\begin{equation}
H^2 = \left(\dfrac{\dot{a}_0}{a_0 n_0}\right)^2 = 
\dfrac{\Lambda_4}{3} - \dfrac{k}{a_0^2} + 
\dfrac{\kappa_4^2}{3}\,(\rho_1 + \rho_2)
\end{equation}
Here the density $\rho = \rho_1 + \rho_2$  appears linearly as in
the usual Friedmann equation. This linear dependence is common to
other induced gravity models in the literature [20] and appears, 
of course with different values of the  quantities involved, in 
all the brane solutions found in this paper. Also we notice 
that in the actual case it is neither possible to recover  the 
brane metric evaluating the bulk metric at $y = 0$ nor in general
to give the explicit dependence of the scale factor on  the 
proper time $\tau$. As a qualitative estimate one can say that  
$a_0^2$, starting from zero, either increases indefinitely or 
oscillates depending on the values of the parameters which appear
in eq. (44).  Exact results can however be obtained if we assume 
$k = 0$, which is a value strongly suggested for our universe 
at the present time.  For a spatially flat universe we have 
from eq. (48) 
\begin{equation} 
a_0 \propto 
\sinh^2{\sqrt{\dfrac{\Lambda_4}{12}}\,\tau}, \quad 
\quad H^2 =  
\dfrac{\Lambda_4}{3}\,\coth^2{\sqrt{\dfrac{\Lambda_4}{12}} 
\,\tau} \quad \mathrm{and} \quad q = -\,\dfrac{1}{2}\ 
\left(1 +\dfrac{\Lambda_4}{3 H^2}\right)
\end{equation}
Recalling that the density parametr $\Omega_\Lambda = 
\Lambda_4/(3\,H^2)$ varies in the interval $[0,1]$, one obtains 
for the deceleration parameter $-\,1 \leq q \leq -\,\frac{1}{2}$ 
and for the age of the universe $\tau_0 = 
2/H_0\,\left(\sqrt{1/(\Omega_\Lambda)_0}\coth^{-1}\sqrt{ 
{1/(\Omega_\Lambda)_0}}\right)$ where the subscript ${}_0$ here
refers to present values. While the range of values for $q$ 
is in agreement with the observational results, this does not 
happen for the age of the universe whose minimum value  here 
results equal to $2/H_0$.\pni Let us finally treat the case which
corresponds to the choice $\phi = 1$ in the bulk. The relevant 
Einstein equations are 
\begin{subequations}
\begin{eqnarray} \hspace{-0.7in}
-\,\dfrac{k}{a_0^2} -\, \dfrac{{\dot{a}_0}^2}{a_0^2 n_0^2} + 
\dfrac{2\,\dot{a}_0\dot{n}_0}{a_0 n_0^3} -\, 
\frac{2\,\ddot{a}_0}{a_0 n_0^2}\, = && \dfrac{2\,\Lambda_5}{3} 
-\, \dfrac{\Lambda_5}{6}\,\left(\dfrac{4}{\sin^2{\dfrac{1}
{\lambda}\,\sqrt{\dfrac{\Lambda_5}{6}}\,t}} +  
\dfrac{1}{\cos^2{\dfrac{1} {\lambda}\, 
\sqrt{\dfrac{\Lambda_5}{6}}\,t}}\right) \nonumber \\ 
&& -\, \dfrac{k}{\dfrac{6\,\alpha^2}{\Lambda_5}\,\sin^2{\dfrac{1}
{\lambda}\,\sqrt{\dfrac{\Lambda_5}{6}}\,t}} \label{equation9} \\
{} \nonumber \\ 
-\,\dfrac{3\,k}{a_0^2} -\, 
\dfrac{3\,\dot{a}_0^2}{a_0^2 n_0^2}\, = && \dfrac{2\,\Lambda_5}{3} 
-\, \dfrac{\Lambda_5}{6}\,\left(\dfrac{6}{\sin^2{\dfrac{1}
{\lambda}\,\sqrt{\dfrac{\Lambda_5}{6}}\,t}} +  
\dfrac{1}{\cos^2{\dfrac{1} {\lambda}\, 
\sqrt{\dfrac{\Lambda_5}{6}}\,t}}\right) \nonumber \\ 
&& -\, 
\dfrac{3\,k}{\dfrac{6\,\alpha^2}{\Lambda_5}\,\sin^2{\dfrac{1} 
{\lambda}\,\sqrt{\dfrac{\Lambda_5}{6}}\,t}} \label{equation10} 
\end{eqnarray}
\end{subequations} 
The solutions to the system (50) are
\begin{eqnarray}
a_0^2 = && c_1\, \exp{\left(-\, \dfrac{\alpha^2}{3\,(k + 
\alpha^2)\, \cos^2{\dfrac{1}{\lambda}\,
\sqrt{\dfrac{-\,\Lambda_4}{4}}\,t}}\right)}\,
\dfrac{\left(\sin^2{\dfrac{1}{\lambda}\,
\sqrt{\dfrac{-\,\Lambda_4}{4}}\,t}\right)^{2-\,k/(k+\alpha^2)}}
{\left(\cos^2{\dfrac{1}{\lambda}\,
\sqrt{\dfrac{-\,\Lambda_4}{4}}\,t}\right)^{\alpha^2/(3\,(k + 
\alpha^2))}} \\ 
n_0^2 = && 
\dfrac{{\dot{a}_0}^2}{\dfrac{a_0^2}{3}\,\left(\Lambda_4 + 
\kappa_4^2\,(\rho_1 + \rho_2 +\rho_3) \right)-\,k} 
\end{eqnarray}
Comparing eqs. (50) and (9) we have defined the 4-D 
cosmological constant as $\Lambda_4 = -\,\dfrac{2\,\Lambda_5}{3}$
and have considered the remaining terms as the superposition of 
three different fluids with pressures and densities given by
\begin{eqnarray}   
p_1 = -\, \dfrac{2}{3}\,\rho_1, \quad \rho_1 = &&-\, 
\dfrac{3\,\Lambda_4} 
{2\,\kappa_4^2\,\sin^2{\dfrac{1}{\lambda}\,\sqrt{\dfrac{-\,\Lambda_4}
{4}}\,t}}   \\
p_2  =  -\,\hspace{0.15in}\rho_2,\quad \rho_2 = &&-\, 
\dfrac{\Lambda_4}{4\,\kappa_4^2\, 
\cos^2{\dfrac{1}{\lambda}\,\sqrt{\dfrac{-\,\Lambda_4}{4}}\,t}} \\
p_3 = -\, \dfrac{1}{3}\,\rho_3,\quad \rho_3 = &&-\, 
\dfrac{3\,k}{\dfrac{4\,\alpha^2}{\Lambda_4}\, 
\kappa_4^2\,\sin^2{\dfrac{1}{\lambda}\,\sqrt{\dfrac{-\,\Lambda_4}{4}}\,t}}
\end{eqnarray} 
Again we notice  that in the actual case it is neither possible 
to recover  the brane metric evaluating the bulk metric at $y = 
0$ nor in general to give the explicit dependence of the scale 
factor on  the proper time $\tau$ even assuming $k = 0$. As a 
qualitative estimate one can say that the scale factor $a_0^2$, 
starting from zero, either increases indefinitely or, after 
increasing, reaches again the zero depending on  the values of 
the parameters appearing in eq. (51). 
\section{C\lowercase{onclusions}}
We have studied brane-world cosmologies where the 
Einstein-Hilbert action is modified by a 4-D scalar curvature 
from induced gravity on the brane. To investigate  cosmological 
solutions in the brane-bulk system, we extended the static 
Randall-Sundrum line element to a simple dynamical case and from 
the knowledge of the metric so obtained we solved the 4-D 
Einstein equations on the brane. The general features of the 
model are presented and  the evolution of physically 
meaningfull quantities can be determined choosing suitably the 
model parameters. In our description the cosmological fluid 
appears as a mixture of perfect fluids obeying simple equations 
of state with constant barotropic factors. Of course our model 
has to be implemented to can describe the  distinct periods 
of the universe evolution, a task which will require to consider 
more complicated equations of state for real fluids. If one 
adopts an approach of this kind, the simple model  described here
may serve as a basis for obtaining more detailed braneworld 
solutions and, therefore, a better comparison with the 
accumulated cosmological observations.


\newpage
\begin{thebibliography}{00} 
\bibitem{1} T. Shiromizu, K. Maeda, M. Sasaki, \textit {Phys. 
Rev.} D 62 (2000) 024012 (gr-qc/9910076);\\ M. Sasaki, T. 
Shiromizu, K. Maeda, \textit{Phys. Rev.} D 62 (2000) 024008 
(hep-th/9912233). 
\bibitem{2} E. Anderson, R. Tavakol, 
\textit{Class. Quant. Grav.}, 20 (2003) L267 (gr-qc/0305013);\\ 
(gr-qc/0309063). 
\bibitem{3} G. Kofinas, \textit{JEHP} 0108 (2001) 034 
(hep-th/0108013). 
\bibitem{4} L. Randall, R. Sundrum, \textit{Phys. Rev. Lett.} 83 
(1999) 4690 (hep-ph/9905221).
\bibitem{5} R. M. Wald, \textit{General Relativity}, The 
University of Chicago Press, 1984. 
\bibitem{6} W. Israel, \textit{Nuovo Cim.} 44 B (1966) 1.
\bibitem{7} G. Kofinas, E. Papantonopoulos, I. Pappa,
\textit{Phys. Rev} D 66 (2002) 104014 (hep-th/0112019) and 
references quoted therein. 
\bibitem{8} D. Langlois, \textit{Prog. Theor. Phys. Suppl.} 148 
(2003) 181 (hep-th/0209261). 
\bibitem{9} C. Grojean, \textit{Phys.
Lett.} B 479 (2000) 273 (hep-th/0002130). 
\bibitem{10} P. Bin\' etruy, J. M. Cline, C. Grojean, 
\textit{Phys. Lett.} B 489 (2000) 403 (hep-th/0007029). 
\bibitem{11} A. Kehagias, K. Tamvakis, \textit{Phys. Lett.} B 515
(2001) 155 (hep-ph/0104195).
\bibitem{12} Ya. B. Zel'dovich, \textit{Uspekhi Akad. Nauk. SSSR}
95 (1968) 209;\\ \textit{Sov. Phys. Uspekhi} 11 (1968) 381. 
\bibitem{13} A. Agnese, M. La Camera, A. Wataghin, \textit{Nuovo 
Cim.} 66 B (1970) 202.
\bibitem{14} H. Liu, P. S. Wesson, \textit{Int. J. Mod. Phys.} D 
3 (1994) 627. 
\bibitem{15} P. S. Wesson, H. Liu, S. S. Seahra, 
\textit{Astron. Astrophys.} 358 (2000) 425 (gr-qc/0003012).
\bibitem{16} G. T. Horowitz, I. Low, A. Zee, \textit{Phys. Rev.}
D 62 (2000) 086005 (hep-th/0004206).
\bibitem{17} J. Ponce de Leon, \textit{Int. J. Mod. Phys.} D 12 
(2003) 1053 (gr-qc/0212036);\\ \textit{Gen. Rel. Grav.} 36 (2004)
923 (gr-qc/0212058).
\bibitem{18} T. Fukui, S. S. Sehara, P. S. Wesson, \textit{J. 
Math. Phys.} 42 (2001) 5195 (gr-qc/0105112).
\bibitem{19} S. S. Seahra, P. S. Wesson, \textit{J. Math. Phys.}
44 (2003) 5664 (gr-qc/0309006).
\bibitem{20} H. Collins, B. Holdom, \textit{Phys.
Rev.} D 62 (2000) 105009 (hep-ph/0003173); E. Kiritsis, N. 
Tetradis, T. N. Tomaras, \textit{JHEP} 03 (2002) 019 
(hep-th/0202037); K. Maeda, S. Mizuno, T. Torii, \textit{Phys. 
Rev.} D 68 (2003) 024033 (gr-qc/0303039); G. Kofinas R. Maartens,
E. Papantonoupolos, \textit{JHEP} 0310 (2003) 066 (hep-th/0307138). 
\end{thebibliography}
\end{document}